\documentclass[12pt, draftclsnofoot, onecolumn]{IEEEtran}

\usepackage{setspace}
\usepackage{cite}

\usepackage[pdftex]{graphicx}
\graphicspath{/figs/}
\DeclareGraphicsExtensions{.pdf,.jpeg,.png}
\usepackage{amsmath, amssymb}

\usepackage[table]{xcolor} 
\usepackage{bm}

\usepackage{float}
\usepackage[caption = false]{subfig}
\usepackage{algorithm}
\usepackage{algorithmicx}
\usepackage{algpseudocode}
\usepackage{pifont}
\usepackage{xcolor}

\usepackage{array}
\usepackage{url}
\usepackage{parskip}
\usepackage{layout}

\let\OLDthebibliography\thebibliography
\renewcommand\thebibliography[1]{
  \OLDthebibliography{#1}
  \setlength{\parskip}{0pt}
  \setlength{\itemsep}{5pt plus 0ex}
}

\usepackage{mathtools}

\algdef{SE}[DOWHILE]{Do}{doWhile}{\algorithmicdo}[1]{\algorithmicwhile\ #1}%

\DeclareMathOperator*{\argmax}{argmax}
\newcommand{\kargmax}[1]{#1\text{-}\argmax}

\newcommand{\ue}{u \color{black}}

\newcommand{\Nx}{N_{\text{X}}}
\newcommand{\Ny}{N_{\text{Y}}}
\newcommand{\Nt}{N_{\text{T}}}
\newcommand{\Nr}{N_{\text{R}}}
\newcommand{\nt}{n_{\text{t}}}
\newcommand{\nr}{n_{\text{r}}}
\newcommand{\Np}{N_{\text{P}}}
\newcommand{\Oh}{O_{\text{H}}}
\newcommand{\Ov}{O_{\text{V}}}

\newcommand{\norm}[1]{\left\lVert#1\right\rVert}

\newcommand{\vander}[2]{\mathbf{a}_{N_{#2}}(#1)}
\newcommand{\vect}[1]{\mathbf{#1}}
\newcommand{\complex}{\mathbb{C}}

\newcommand{\Hfu}[1]{\vect{H}_{\ue, #1}}

\newcommand{\Hstack}{\vect{H}}
\newcommand{\Htkstack}[1]{\vect{H}_{#1, t, k}}

\newcommand{\HFest}{\widehat{\vect{H}\vect{F}}}

\newcommand{\Hbsu}{\widehat{\widehat{\vect{H}}}_{\ue}}
\newcommand{\Hbs}{\widehat{\widehat{\vect{H}}}}

\newcommand{\sinr}{\text{SINR}}

\newcommand{\Kdmrs}{K_{\text{DMRS}}}
\newcommand{\Tdmrs}[1]{\mathbf{T}_{{#1},\text{DMRS}}}
\newcommand{\Kssb}{K_{\text{SSB}}}
\newcommand{\Tssb}[1]{T_{{#1},\text{SSB}}}

\newcommand{\Pcsi}{P_{\text{CSI}}}

\newcommand{\F}{\mathbf{F}}
\newcommand{\f}{\mathbf{f}}
\newcommand{\Lmax}{L_{\text{{max}}}}
\newcommand{\Lcsi}{L_{\text{{CSI}}}}
\newcommand{\Spilot}{\vect{s}}

\newcommand{\RSRP}{\text{RSRP}}
\newcommand{\RSRPiu}{\usel{\RSRP_i}}

\newcommand{\Fssb}{\mathbf{F}_{\text{SSB}}}
\newcommand{\Fcsi}{\mathbf{F}_{\text{CSI-RS}}}
\newcommand{\Bssb}{\mathbf{B}_{\text{SSB}}}
\newcommand{\Bcsi}{\mathbf{B}_{\text{CSI-RS}}}
\newcommand{\Bdft}{\mathbf{B}_{\text{DFT}}}

\newcommand{\q}[1]{\mathbf{Q}_{#1}}
\newcommand{\Cq}{\vect{A}}
\newcommand{\bwp}{\text{BWP}}

\newcommand{\usel}[1]{#1^{(\ue)}}

\newcommand{\topic}[1]{{#1}}

\begin{document}

\title{ML Codebook Design for Initial Access and CSI Type-{II} Feedback in Sub-6GHz {5G NR}}
\author{\IEEEauthorblockN{Ryan M. Dreifuerst,~\IEEEmembership{Student Member,~IEEE, and }}%
    \and
	\IEEEauthorblockN{Robert~W.~Heath~Jr.~\IEEEmembership{Fellow,~IEEE}}%

	\thanks{Ryan M. Dreifuerst and Robert W. Heath Jr. are with North Carolina State University, Raleigh, NC 27695 \{rmdreifu, rwheathjr\}@ncsu.edu.
	    This material is based upon work supported by the National Science Foundation under grant nos. NSF-ECCS-2153698, NSF-CCF-2225555, NSF-CNS-2147955 and is supported in part by funds from federal agency and industry partners as specified in the Resilient \& Intelligent NextG Systems (RINGS) program.}
}

\maketitle
\bstctlcite{IEEEexample:BSTcontrol}

\vspace{-10pt}
\begin{abstract}
    Beam codebooks are a recent feature to enable high dimension multiple-input multiple-output (MIMO) in 5G new radio (NR). Codebooks comprised of customizable beamforming weights can be used to transmit reference signals and aid the channel state information (CSI) acquisition process. Codebooks are also used for quantizing feedback following CSI acquisition. In this paper, we characterize the role of each codebook used during the beam management process and design a neural network to find codebooks that improve overall system performance. Evaluating a codebook is not purely about maximizing signal power. Instead, it is about considering the system-level dependency between the codebooks, feedback, overhead, and spectral efficiency. The proposed neural network is built on translating codebook and feedback knowledge into a consistent beamspace basis similar to a virtual channel model. Then, the beamspace is fed into the neural network to generate initial access codebooks. This beamspace codebook algorithm is designed to directly integrate with current 5G beam management standards (Releases 15-17), without changing the feedback format, beam training timing, or requiring additional side information. Our simulations show that the neural network codebooks improve over traditional codebooks in received signal power, even in dispersive sub-6GHz environments. We further use our simulation framework to evaluate type-II CSI feedback formats with regard to effective multi-user spectral efficiency. Our results suggest that optimizing codebook performance can provide valuable spectral efficiency improvements, but optimizing the feedback configuration can be equally important for multi-user performance in sub-6GHz bands.
\end{abstract}

\section{Introduction}
    Fifth-generation (5G) cellular systems have adopted a beam-based approach for MIMO communications. In such systems, broadcast control signals are beamformed to provide array gain \cite{Dreifuerst2023magazine}. This helps increase coverage and meet link budgets at high carrier frequencies. 
    While the motivation for beam-based systems was to provide array gain in millimeter-wave (mmWave) frequency bands, the same framework is also used at low frequencies. Unfortunately, optimization of the beamforming framework, and subsequent feedback strategies, has not been extensively investigated for sub-6GHz multi-user MIMO (MU-MIMO), despite the importance of low- and mid-band MIMO in current and anticipated future standards \cite{Lee2022XMIMO}.

    \topic{Codebook-based beam management is a framework for beamforming and feedback that enables transmitting beamformed reference signals and multiple forms of CSI acquisition \cite{Dreifuerst2023magazine}}. In early 3GPP releases up to Release $12$, channel state information was obtained by estimating the channel from reference signals for every transmitter-receiver beam pair using unprecoded signals. Unprecoded reference signals were used up through 4G LTE Release $14$ because multi-antenna beamforming was not needed to achieve higher signal-to-noise ratio (SNR) for synchronization and channel estimation \cite{Morozov2016}. Beamformed control signals have become a necessity in mmWave bands as a result of the increasing bandwidths (which reduce the SNR for a constant transmit power)
    and shrinking antennas sizes--resulting in pre-beamformed SNR on the order of $-15$dB or less \cite{Raghavan2016BFIA}.
    By beamforming the control and reference signals, receivers are more likely to detect initial access signals and obtain more accurate synchronization and channel measurements.

    The process of obtaining CSI at the transmitter involves multiple steps and codebooks to balance the overhead, latency, and accuracy of the CSI \cite{HengSixChallengesBM6G2021, Giordani3GPPBeamManagement2019}. To begin the process, the base station (BS) transmits a synchronization signal block (SSB) with a beamformer selected from a codebook (SSB codebook) that enables synchronization and random access. The SSB codebook is typically small and all of the codewords in the codebook are used during each SSB period. The user equipment (UE) will provide a small amount of feedback to conclude the initial access that includes a beam selection from the SSB codebook. The BS will use this information to select a subset of beamformers from a large codebook, different from the SSB codebook, for transmitting channel state information reference signals (CSI-RS). We assign this codebook the name ``CSI-RS codebook''. The relationship between the SSB and CSI-RS codebook can be understood from the perspective of a hierarchical beam search \cite{Alkhateeb2014}. The CSI-RS enable hybrid beam training by providing multiple precise beams for analog beam training--often referred to as beam refinement--and pilots for digital channel estimation. The UE will provide an in-depth packet of feedback following CSI-RS transmission that includes the CSI-RS codebook beamformer selection as well as information based on the channel estimation. While the first iterations of beam management used wide and narrow discrete Fourier transform (DFT) beamformers \cite{Raghavan2016BFIA}, there is no requirement that the beamformers be ``wide'' and ``narrow'' for the SSB and CSI-RS codebooks. In fact, the codebooks can be arbitrary since the UEs do not need knowledge of the beamformers during initial access, beam refinement, or even data transmission.

    While the SSB and CSI-RS codebooks can be arbitrary, the third codebook used for quantizing the feedback (denoted as the FB codebook) is precisely defined depending on the format (type-I, type-II). The FB codebook is defined for specific BS and UE array geometries so that the CSI can be represented and fed back as indices from the codebook. This codebook must be known at the BS and UE.
    It is not strictly necessary that even the feedback codebook is used when reciprocity exists, as reciprocity between the uplink and downlink channel can be exploited in time-division-duplexing systems. Feedback from downlink training, though, is still beneficial in these systems due to asynchronous link budgets \cite{Bjornsson2022CombiningCSIFeedback} or when the number of transmitting and receiving antennas on a UE is not equal \cite{Samsung2020}.

    Our work focuses on a realistic and system-level evaluation of the codebooks used during the beam management framework in sub-6GHz bands. In particular, we unify the two stages of codebook design with the feedback framework and MU-MIMO data transmission to achieve good network performance. Our contributions in this manuscript are summarized as follows:
	   \begin{itemize}
            \item First, we precisely describe the beam management and CSI type-II feedback approach for 5G networks. There are many misconceptions on how the codebooks, SSB, and CSI-RS processes are used for obtaining different forms of feedback. We explain how the various codebooks are used, what is specified in the standards versus left up to implementations, and provide an efficient implementation for feedback quantization in CSI type-II.
            
            \item Second, we present a novel neural network architecture, Beamspace-Codex (BSC), and feedback processing technique for SSB codebook generation. We constrain the system to follow 5G beam management timing which presents a significant constraint for dynamic SSB and CSI-RS codebook generation. To mitigate this constraint, the SSB codebook is dynamically generated--which is manageable at the SSB periodicity--and the CSI-RS codebook is determined as a precomputed DFT decomposition of the SSB codebook. With this integration, we unify the arbitrary codebooks learned via deep learning with the feedback quantization technique to facilitate a hierarchical beam search that improves the beam management process used in 5G NR.
            
            \item Third, we present the results of an extensive evaluation of our proposed method with benchmark DFT codebooks and CSI-based eigen-beamforming solutions using various levels of feedback and overhead. We show that our proposed solution achieves significant performance improvements. We also show that MU-MIMO performance is dramatically limited by the quantization resolution techniques in low-band 5G. In contrast, the frequency selectivity and resource allocation have little impact in the current framework and scattering environment.
        \end{itemize}   

    There is significant work on beam training and beam alignment. Here we review a set of relevant prior work relating to machine learning and beam training. In \cite{LiBeamTraining2013}, a gradient-optimization approach, not using deep learning, is shown to be successful in aiding the beam search process with respect to maximizing the SNR in $60$GHz personal networks. Another approach using classical data-driven techniques is presented in \cite{V.VaEtAlInverseMultipathFingerprintingMillimeter2018} for vehicular networks based on historical beam training results. While the codebooks and algorithms are relatively straightforward, the idea of ``learning'' an efficient codebook from site-specific data is a cornerstone of many modern beam alignment techniques \cite{WangEtAlSiteSpecificCompressiveCodebook2021, Xue2022BMFRL, Yang2022DLBeamAlignment}. These papers \cite{LiBeamTraining2013, V.VaEtAlInverseMultipathFingerprintingMillimeter2018, WangEtAlSiteSpecificCompressiveCodebook2021, Xue2022BMFRL, Yang2022DLBeamAlignment} all focus on mmWave channels and often exploit the sparsity of such environments to aid the beam training process. Furthermore, all of these papers focus exclusively on beam alignment with standard orthogonal codebooks like DFT and phase-shifted codebooks. While DFT codebooks are sensible for certain array geometries and limited scattering environments at mmWave bands, these codebooks do not necessarily work well in rich scattering. Furthermore, none of these papers consider a MIMO format with multiple UE antennas and streams of data. In this paper, we consider a full MU-MIMO OFDM system in a rich scattering environment and use dynamic codebooks.

    In another line of related work, algorithms have been proposed that design beamformers or codebooks \cite{OrigHierarchical2009, Alkhateeb2014, XiaoCodebook2016, ShafinEtAlDRL4BeamOptim2020, XiaMISOBFwithDL2020, Xue2022BMFRL, HengProbingBeams2021, Dreifuerst2022VTCBM}. Hierarchical codebooks and the associated design process were proposed in \cite{OrigHierarchical2009, Alkhateeb2014, XiaoCodebook2016} which separated the beam training into multiple stages of wide-beam and narrow-beam codebooks. A similar strategy was adopted by 5G using the SSB and CSI-RS codebooks and plays an important role in systems designed for 5G deployment. Deep learning was applied to the task in \cite{ShafinEtAlDRL4BeamOptim2020}, which uses deep reinforcement learning to design broadcast beam patterns. While designed for small-scale MIMO, the results suggest that arbitrary learned beam patterns can be an effective way to initiate communication in broadband networks. A supervised and an unsupervised learning approach were presented in \cite{XiaMISOBFwithDL2020}, which showed promising results when used to design beamforming vectors from channel measurements. These results were based on a simplified channel model, though, and perfect CSI was assumed at the BS. Recently, a deep, federated, reinforcement learning strategy was proposed \cite{Xue2022BMFRL} that jointly trained a beam management model over a network using user location information. The focus was on deciding which sectors would be active, thereby narrowing the ``codebook,'' although the paper ignores all beamforming besides sector-level association and large-scale pathloss. Another paper considered the task of learning probing beams \cite{HengProbingBeams2021}, which is one of the only papers to consider both the SSB and CSI-RS codebooks. The results show significant gains can be achieved over a basic hierarchical search, although perfect CSI is assumed to be available for training and the channels are assumed to be narrowband. The results from \cite{HengProbingBeams2021} provide motivation for researching realistic deep learning codebook methods. In our previous work we designed a neural network for SSB codebook generation in narrowband channels as well but trained using an angular representation of the channel \cite{Dreifuerst2022VTCBM}. The results, however, did not extend well to wideband systems and did not consider the impact that the SSB codebook has on system-level performance. In this investigation, we now model wideband channels and fully incorporate the entire beam management and data transmission into our evaluation.
    
    An important part of the 5G beam management framework is the connection between beam training and feedback. An enhanced CSI feedback strategy was proposed for full-dimension MIMO \cite{Morozov2016} in 4G Release $13$ which is based on a linear combination of DFT codewords. This strategy is very similar to the type-II feedback format introduced in 5G Release $15$. One key difference is that the initial idea assumed the users must know the beamforming vector used during pilot transmission, which was not adopted into the standards. There has also been growing interest in using machine learning to design or modify the feedback \cite{WenDL4mMIMOCSIFeedback2018, KimLearningMISOBF2021, Chen2021}. All that work assumes that both the network and the users can share models for encoding and decoding the feedback, which is not supported or easily introduced into network operation. In this paper, we do not use deep learning to change the feedback methods. Instead, we characterize the relationship and limitations the feedback and beam management strategy have on MU-MIMO performance.
    
	\begin{figure}[ht!]
	    \centering
	    \includegraphics[width=6in]{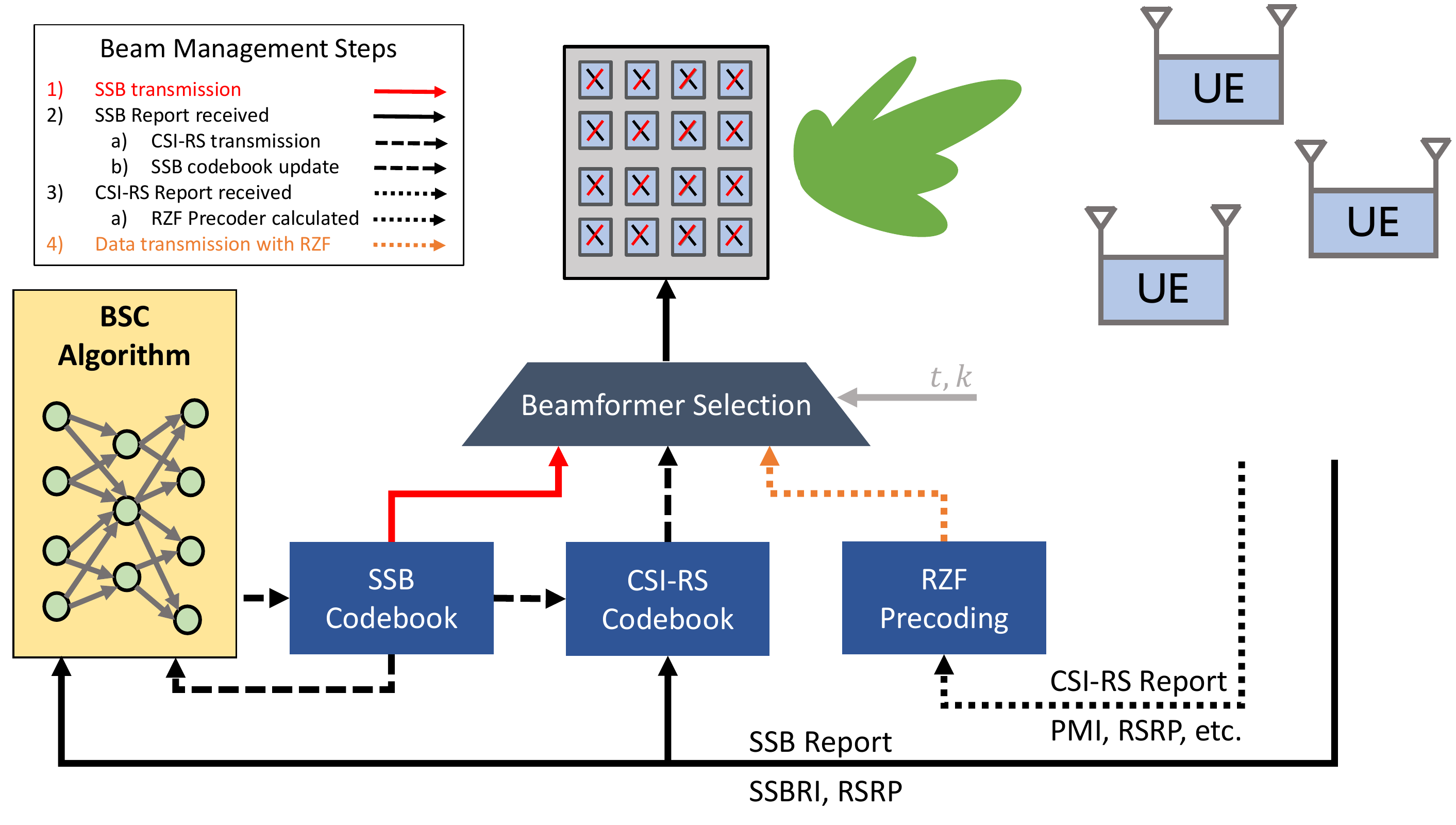} 
	    \caption{Beamforming codebooks are used to transmit reference signals for synchronization from the SSB codebook. Based on the SSB report, additional reference signals are beamformed using the CSI-RS codebook for channel estimation by the user equipment (UEs). Using the CSI-RS report, the base station can design a precoder (i.e. RZF precoder) that attempts to maximize the spectral efficiency. The proposed BSC algorithm updates the SSB codebook, which in turn affects the CSI-RS codebook and feedback.}
	    \label{fig: system_image}
	\end{figure}

    $\mathbf{Notation}$: $\vect{A}$ is a matrix, $\vect{a}$ and $\{a[i]\}_{i=1}^{N}$ are column vectors and $a,A$ denote scalars. $\vect{A}^T$, $\overline{\vect{A}}$, $\vect{A}^*$, and $\vect{A}^{\dagger}$ represent the transpose, conjugate, conjugate transpose, and psuedo-inverse of $\vect{A}$. 
    $\vect{A}[k, \ell]$ denotes the entry of $\vect{A}$ in the $k^{\text{th}}$ row and the $\ell^{\text{th}}$ column. The same meaning is also associated with $\vect{A}_{k, \ell}$. Similarly, $\vect{A}[:, k]$ refers to the $k^\text{th}$ column of $\vect{A}$. Unspecified norm equations are $\norm{\mathbf{a}}_{2} = \mathbf{a}^* \mathbf{a}$ for vectors and the Frobenius norm $\norm{\mathbf{A}}_F = \sqrt{\text{Tr}(\mathbf{A} \vect{A}^*)}$ for matrices. We make use of the disjoint operator $\vect{A}\ \backslash\ \boldsymbol{\alpha}$ to describe the matrix (or vector) $\vect{A}$ excluding the subset (or element) $\boldsymbol{\alpha}$. We define $\mathsf{j}=\sqrt{-1}$. The operator $\mathbb{E}[\cdot]$ is used for the expectation of a random variable. Due to the notational complexity of MU-MIMO with OFDM, we will always use $u$ to refer to a specific UE, $t$ as a specific time, $k$ as a specific frequency resource, and $\nt / \nr$ to refer to a specific transmit or receive antenna. 

    The remainder of our paper is organized as follows. First, we introduce the system model beginning from an arbitrary MU-MIMO OFDM channel and continuing through the SSB and CSI-RS beam management processes. We finish this section with a description of the channel estimation, feedback, and data transmission stages. Next, we introduce the beamspace observation and neural network architecture that form the basis of our work. In the final sections, we present the simulation setup and evaluate the various codebooks and feedback formats to address the lack of prior work when it comes to codebooks and CSI formats in sub-6GHz 5G NR.

\section{System model} \label{sec: system}
    We begin the investigation with an overview of the channel definition and assumptions before introducing the beam management steps including SSB transmission, CSI-RS transmission, feedback, and downlink data transmission, as shown in Figure \ref{fig: system_image}. Afterward, we outline the problem formulation that influences the neural architecture. Throughout this paper, we will limit the system configuration to a single cell with multiple UEs each equipped with multiple antennas and all arrays are fully digital and single-polarization. While the inclusion of multiple cells is more realistic, the impact of multi-cell interference on beam management and feedback can be mitigated with limited coordination between nearby cells. The use of single-polarized arrays is also a simplification, although during beam training the second polarization operates as a simple extension with little impact on the overall system operation. We will investigate multiple BS deployments with dual-polarized arrays in future work.

    \subsection{MU-MIMO channel model}
        We model the system so that it is representative of real-world conditions to evaluate performance in a realistic beam training scenario for sub-6GHz massive MIMO. We do not explicitly specify TDD or FDD because this work does not rely on channel reciprocity. In addition, the channel will always be handled in the frequency domain as the multi-user downlink OFDM channel, however, we do not assume any specific channel model at this point. The OFDM MU-MIMO channel is defined for $U$ users, over $T$ time slots, and $K$ subcarriers between each of the $\Nr$ receive antennas and $\Nt$ transmit antennas as $\Hstack \in \complex^{U \times T \times K \times \Nr \times \Nt}$.
        Note that we can generate this channel model using a clustered channel model as typically done by QuaDRiGa \cite{JaeckelQuadriga}, 3GPP, and other raytracing or statistical models with the array response vectors $  \vander{\theta}{}=\frac{1}{\sqrt{N}}[1, e^{j \pi \cos{\theta}}, e^{j 2 \pi \cos{\theta}}, ...,\ e^{j(N-1) \pi \cos{\theta}}]^T$.
        Defining the time and frequency responses for every antenna pair is critical to accurately model the beam training process with specific resource elements allocated for pilots (reference signals) and data. With this setup in mind, a generic received signal model for signal $\vect{s}$ transmitted with precoder/beamformer $\vect{F}$ and received with a combiner $\vect{W}$ over a band of subcarriers $K$ is 
        \begin{align}
            \vect{y}_{t, k}^{(\ue)} &= \frac{1}{\sqrt{K\Nt}} \vect{W}_{t, k}^{(\ue)} \Hfu{t, k} \sum_{\text{u}=0}^{U-1} \vect{F}_{\text{u}, t, k} \vect{s}_{\text{u}, t, k} + \vect{W}_{t, k}^{(\ue)}\vect{N}_{t, k}. \label{eqn: rec_signal}
        \end{align}
        The noise $\vect{N}$ is modeled as IID, complex Gaussian random values to account for thermal noise and the noise figure of the receiver. Note that because users do not share information, the combiners and received signals have superscript $(\ue)$, while the beamformers and channels have subscript $\ue$ to imply that an aggregation is possible i.e. the BS has knowledge of the precoders for all users.
        During the next subsections, we will specify how $\usel{\vect{W}}, \vect{F}_{\ue},$ and $\vect{s}_{\ue}$ are used during the SSB, CSI-RS, and data transmission stages. We will assume throughout the paper that all beamformers are normalized according to a per-symbol power constraint, i.e. $\norm{\vect{F}}^2 =\ \Nt$. In the SSB and CSI-RS stages, $\vect{F}$ is selected from one or more codebooks that play critical roles in beam management.

    \subsection{Codebooks}
        Communication protocols employ codebooks in many different ways for beam training, feedback representations, information quantization, and more. Generally, codebooks are understood as an ordered set of elements called codewords. Throughout this work on beam management, codewords will correspond to a vector of complex values and we will treat codebooks as matrices to simplify notation in later steps. There are three codebooks that we will reference and directly impact beam management: a codebook of beamforming vectors for SSB transmission (SSB codebook), a codebook for beamforming CSI-RS transmissions (CSI-RS codebook), and a codebook for feedback quantization (FB codebook). Each of these codebooks is designed to serve different purposes, although there are many inter-relationships that make designing and evaluating codebooks challenging.
        
        The sizes of the codebooks are based on the supported usage for each one. The SSB codebook $\Bssb \in \complex^{\Np \times \Lmax}$ has $\Lmax$ codewords which is between $1$ and $64$ depending on the carrier frequency \cite{Dreifuerst2023magazine}. The CSI-RS codebook $\Bcsi \in \complex^{\Np \times N}$ is at least the same size, where $N$ is typically large but only a small subset of beamformers is selected to be used in one period. The FB codebook $\Bdft \in \complex^{\Np \times \Nx \Oh \Ny \Ov}$ is an oversampled DFT codebook with horizontal and vertical oversampling factors $\Oh$ and $\Ov$. The number of logical ports, $\Np$, determines the digital precoding dimension and is equal to $\Nx \Ny$ in a fully digital array. The first two codebooks are transparent to the UE, meaning that they are configurable and any arbitrary beamforming codebooks can be employed. The FB codebook must stay constant and is fully determined by the BS logical array size and oversampling factors, which are sent to the UE during initial access so that both the BS and the UE can use the same codebook for feedback representation. Previously, 4G used a feedback codebook based upon predefined matrices called CSI type-I precoding matrix indicator (PMI). 5G augmented the feedback to include type-II PMI which corresponds to a set of directional array responses with amplitude and phase combining factors.

    \subsection{SSB transmission and reception}
        \topic{Initial access is used in mobile networks to obtain limited channel information, achieve synchronization, and set up random access procedures.} In 5G NR, a cell may initiate the initial access period at regular intervals of $\{5, 10, 20, 40, 80, 160\}$ms \cite{Giordani3GPPBeamManagement2019}, to control the periodicity a UE must be active (not in power-saving mode) and provide feedback to the network. During this period, the cell will transmit SSBs that contain primary and secondary synchronization signals (PSS, SSS) as well as demodulation reference signals (DMRS) \cite{Giordani3GPPBeamManagement2019}. These SSBs are beamformed using a specific beamforming vector selected from $\Bssb$ and associated with a beam index. Each SSB is transmitted sequentially in time, using $20$ contiguous resource blocks. This is a much smaller bandwidth than the downlink data transmission and is a key limitation of relying on SSB feedback to obtain CSI in frequency-selective channels. Furthermore, the resource allocation of SSBs is essentially fixed to assist new users with joining a network. Depending on the cell carrier frequency and subcarrier spacing, a cell may transmit up to $\Lmax = \{1, 4, 8, 64\}$ beams in a period, and all cells must transmit at least one beam. During transmission, the UE will measure the Reference Signal Received Power (RSRP) and report the measurement using the random access channel slot corresponding to the index of the beam with the highest RSRP. This way the base station knows how strong the signal was and to which of the SSB codebook's codewords it corresponds. 

        The SSB is transmitted over a band of $\Kssb$ frequency resources at times $\Tssb{i}$ for the $i^{\text{th}}$ SSB using beamformer $\f_i \in \Bssb$. While transmitting the reference signal, the rest of the band is still available for data transmission, so the system power is shared over all $K$ resource blocks. In total, all $\Lmax$ SSBs are transmitted from the codebook $\Bssb$. 
        During SSB reception, the generic received signal model from \eqref{eqn: rec_signal} is modified to account for the single data stream
        \begin{align}
            y_{ \text{SSB}_i, t, k}^{(\ue)} = \frac{1}{\sqrt{K \Nt}} \vect{w}_{t, k}^{(\ue) *}  \Hfu{t, k} \f_i s_{t, k} + \vect{w}_{t, k}^{(\ue) *} \vect{N}_{t, k}. \label{eqn: rec_sig_2}
        \end{align}
        In mmWave bands, it is common that $\vect{w}$ is also determined through beam training. In sub-6GHz bands, the receiver instead uses antenna selection or simply the first antenna for power saving \cite{3gppTS38.214}.
        With the received signal model \eqref{eqn: rec_sig_2} in mind, we can now define an important quantity provided as feedback during initial access, the RSRP.
 
    \subsection{Reference signal receive power}
        RSRP is one of the primary metrics that the receiver will measure during initial access for determining the channel quality. The base station will use the RSRP feedback to determine the strength of the signal, the initial code rate to be used, and what CSI-RS beamformers should be employed at the next iteration. The RSRP measures the received power during a given reference signal (SSB or CSI-RS). In the SSB case, the demodulation reference signal (DMRS), known at the transmitter and receiver, is a pilot signal used for RSRP measurement and decoding the master information block provided in the SSB. 
        We will assume antenna selection for the $\nr \in \{0, .. \Nr-1\}$ antenna during SSB reception, but because it is fully digital the UE can determine the antenna selection after receiving the full SSB and selecting the strongest receive branch. The RSRP is then
        \begin{align}
            \vect{y}_{ \text{SSB}_i, t, k}^{(\ue)} &=\ \frac{1}{\sqrt{K \Nt}} \Hfu{t, k} \f_i s_{t, k} + \vect{N}_{t, k} \\
            \RSRPiu &=\ \max_{\nr} \sum_{k\in \Kdmrs} \sum_{t \in \Tdmrs{i}} \norm{\vect{y}_{ \text{SSB}_i, t, k, \nr}^{(\ue)}}^2. \label{eqn: RSRP}
        \end{align}
        Note that \eqref{eqn: RSRP} is for a specific UE $(\ue)$ and SSB $(i)$, so the total SSB information aggregated at the base station is comprised of the strongest RSRP and corresponding SSB index (SSBRI) for each of the UEs, denoted as $(\vect{p}, \vect{m})$, and defined as
        \begin{align}
            \vect{p} &=\ \left\{\max_i{\RSRPiu} \right\}^{U-1}_{\ue=0} \label{eqn: Prsrp} \\
            \vect{m} &=\ \left\{\argmax_i{\RSRPiu} \right\}^{U-1}_{\ue=0}. \label{eqn: Irsrp}
        \end{align}
        It is worth noting that the actual SSBRI is not transmitted by the users; the random access slot that the UE responds during will correspond to a specific SSBRI. This is an efficient way to reduce the amount of feedback, allow the base station to perform receive combining for the UE feedback, and still share the necessary information.
        
    	\begin{figure*}[!t]
		    \centering
		    \includegraphics[width=6.25in]{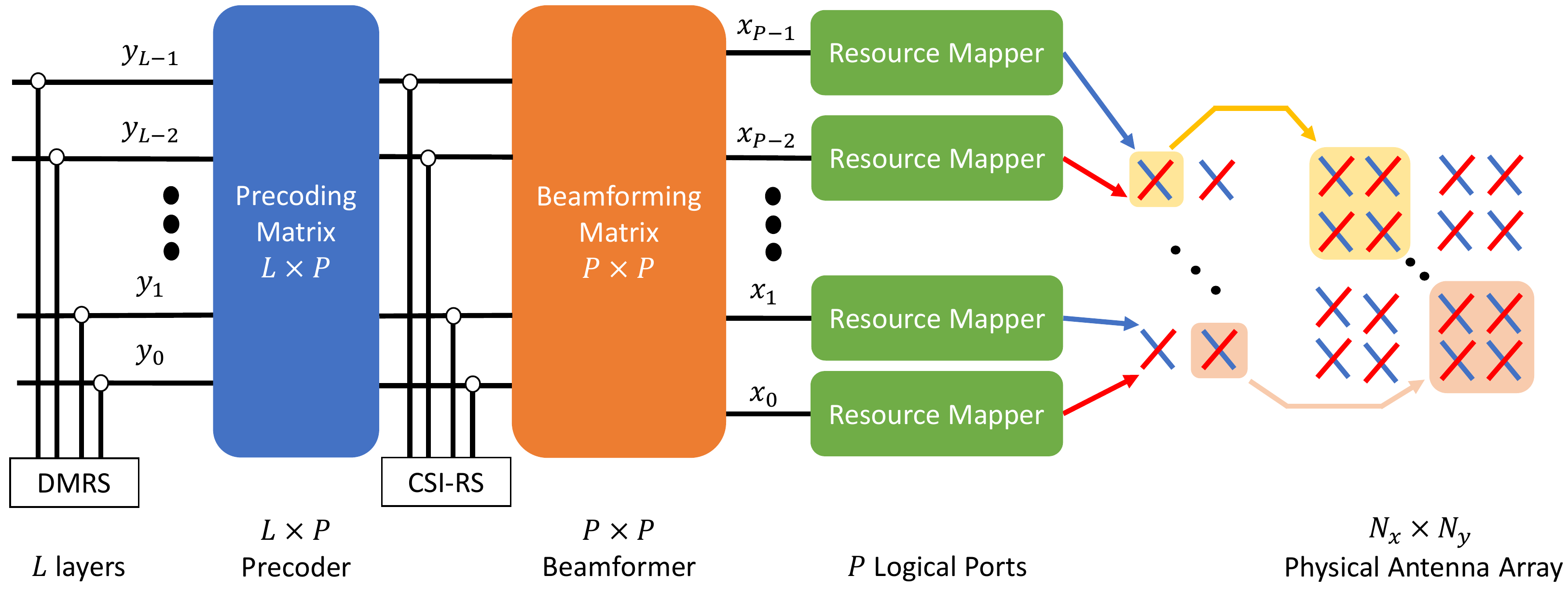}
		    \caption{The data processing flow from data streams (layers) to antenna outputs in 5G. First, the streams (and DMRS) are multiplexed according to the precoding matrix, which essentially assigns the layers to the ports. Then, the result, along with CSI-RS, is beamformed to the logical ports. The logical processes are then mapped onto OFDM resource grids and potentially beamformed again onto the corresponding physical antennas.}
		    \label{fig: ports}
		\end{figure*}
		
    \subsection{CSI-RS transmission and reception}
    	\topic{Beam refinement is the second step of the hierarchical beam management process that is used to obtain better beam alignment and enable wideband channel estimation.} Beam refinement is done through the transmission of CSI-RS which is an encompassing term used for the reference pilots, the beamformed reference signal, and the logical ports that the pilots are sent on. In contrast to SSB signals, CSI-RS are extremely flexible with many combinations of time and frequency resources allocatable. There are some limitations on how often CSI-RS can be transmitted, but the default periodicity a UE expects when joining the network is $80$ms and the symbols must span a larger bandwidth than the SSB--at least $24$ resource blocks and up to the entire bandwidth should contain CSI-RS symbols for a measurement report. Furthermore, CSI-RS are also transmitted both periodically and aperiodically to assist beam tracking with highly mobile users. All told, the beam refinement stage requires more resources than initial access, which is why a hierarchical beam search is used to limit the number of CSI-RS processes needed to achieve a strong connection. 

        Once a BS receives the SSB feedback for a new user, it can attempt to gather more precise CSI through channel state information reference signals. The CSI-RS (and associated CSI-RS beamformers $\Fcsi$) are used for refined channel estimation, although the estimation is performed on the beamformed channel $\vect{H}_{\ue}\Fcsi$, rather than the physical channel. This enables larger hybrid arrays and improves the SNR for more accurate estimation. Readers are encouraged to review \cite{Giordani3GPPBeamManagement2019, HengSixChallengesBM6G2021, Samsung2020} for more details of CSI-RS. The BS can use the SSB feedback to determine and transmit a set of CSI-RS signals with beamformers that are intended to further increase the RSRP compared to the SSB. 
        
        The design of CSI-RS codebooks has been studied to some degree, but DFT beams are common \cite{Morozov2016, Xue2022BMFRL, Singh2021SurveyHybridBF} (and references therein). 
        Often, it is assumed that the set of beamformers is selected from the same codebook as the codebook used for feedback, which \textit{is} specified as an oversampled DFT codebook. There is no specification or need for the SSB or CSI-RS codebooks to be DFT codebooks, although using DFT beamformers during CSI-RS has advantages related to fast and efficient processing. For example, the CSI-RS beamformers can be efficiently selected as the largest DFT components of the SSB beamformers. We will assume the CSI-RS codebook, $\Bcsi$, is comprised of $\Nx\Ny\Oh\Ov$ oversampled DFT vectors throughout this paper, although additional investigations into joint SSB--CSI-RS codebook algorithms are planned in future work. The selection of the CSI-RS beams from SSB decomposition is further refined in Section \ref{sec: algorithm}.
        
         The CSI-RS process involves transmitting a set of pilots $\Spilot$ on the $N_{\text{P}}$ ports along with $\Pcsi$ beamformers $\Fcsi = \{\vect{f}_i\}_{i=0}^{\Pcsi} \subseteq \Bcsi$ chosen by the serving base station to improve the UE channel estimation SNR for the effective channel. The beamformers are used during non-overlapping time-frequency resources ($\vect{T}_{i, \text{CSI}}, \vect{K}_{i, \text{CSI}}$), with the aggregate set of beamformers corresponding to what we define as $\Fcsi$. Each CSI-RS allocation will have at least $\Pcsi$ OFDM resource elements allocated for channel estimation in one resource block and only one CSI-RS beam is active for a given resource to allow for a similar, interference-free beam selection as the SSB process. This is done by setting all other CSI-RS resources and ports to be zero-power CSI-RS except the active one for the specified resources. In a dual-polarized system, the two polarizations are transmitted together to estimate a co-polarization factor that would be fed back during CSI reporting. The whole CSI-RS process is allocated over $\text{NRB}$ resource blocks, which is a configurable parameter to use wider bandwidths and noise averaging during channel estimation. The beam reception and selection process is identical to the RSRP and SSB selection in the previous subsection. For more information on the CSI-RS beam and port allocation please see \cite{Morozov2016} and references therein. At this stage, the entire UE antenna array is active and the UE can perform maximum ratio combining $\usel{\vect{w}_{i, t, k}} = \left(\widehat{\vect{H}_{\ue, t, k}\vect{f}_i} \right)^*$ based on the estimated beamformed channel and appropriate normalization. The estimated SNR at the receiver is
         \begin{equation}
             \text{SNR}^{(\ue)} = \max_{c \in \Pcsi} \frac{1}{K} \sum_{k \in \vect{K}_{c, \text{CSI}}} \sum_{t \in \vect{T}_{c, \text{CSI}}} \frac{1}{N_T} \frac{\norm{\widehat{\vect{H}_{\ue, t, k}\vect{f}_c}}^2}{\mathbb{E}[\norm{\vect{N}_{t, k}}^2]}. \label{eqn: SNR} 
         \end{equation}
         Note that the noise power $\mathbb{E}[\norm{\vect{N}_{t, k}}^2]$ can be estimated using zero-power CSI-RS resources.
         The SNR can be reported during feedback, although one of the most important aspects of the SNR is how the channel estimation (and subsequent feedback) is impacted. For the next part of this section, we will focus on a single user ($\ue$) with SNR corresponding to the channel estimation SNR for the beamformed channel $(\vect{H}_{\ue, t, k}\Fcsi)$. 

    \subsection{Channel estimation}
        \topic{Channel estimation accuracy depends on the number of pilot symbols and the received SNR of the pilots.}
        The algorithm for UE channel estimation is not defined in the specification and is not critical for our system-level analysis. 
        For simplicity, we will assume channel estimation via a simple least squares algorithm so that only knowledge of the pilots is required. Now, we motivate the inclusion of channel estimation in beam management by considering the relationship between the channel estimation error, the codebook-dependent-SNR, and the number of pilot resources $N_{\text{pilots}}$. The resulting mean squared error (MSE) of the channel estimate is \cite[section 3.7]{HeathLozano2018}
        \begin{equation}
            \text{MSE} = \frac{1}{\frac{N_{\text{pilots}}}{N_T} \text{SNR}}. \label{eqn: mse}
        \end{equation}
        Therefore, we can see that a primary feature of the beam management process is to enable more accurate channel estimation by beamforming the reference pilot signals. It might seem that the overhead of using multiple beams is wasted given that the number of pilots could alternatively be increased according to \eqref{eqn: mse}. The issue with such logic is that: 1) it assumes that the signal is always detectable over the noise floor, which is not necessarily the case for non-beamformed signals and 2) the beamforming can provide a gain on the order of $20$dB or more, yet that would require $100$ times more pilot symbols, far larger than the number of beams used.
        
        Calculating the estimated effective channel, $\HFest$, which may not be frequency selective, 
        is the first step in determining the precoder matrix information (PMI), which is one of the quantized CSI elements provided during feedback. The PMI can correspond to different aspects of the estimated channel because it is left to operator implementation as part of the PMI selection procedure. For example, the user could provide PMI according to its desired beamformers (calculated via singular value decomposition of $\HFest$). The base station could then use this information to beamform to a given user and null steer \cite{Godara1997NullForming} the beams for other users. 
        This puts the majority of the computational burden, however, on the UE. Instead, we follow models similar to \cite{Castellanos2018ReconstructHybridPrecoding} where the user provides PMI corresponding to the effective channel estimate, rather than the SVD of the effective channel estimate. This way the BS handles most of the computation, channel reconstruction, and precoder determination.

    \subsection{CSI feedback and reconstruction}
        \topic{We will focus on CSI type-II feedback (Release $16$), as our feedback baseline, which is intended for MU-MIMO with up to $2$ layers per user \cite{3gppTS38.214}, and is also compliant with Release $17$.}
        The UE quantizes the channel estimate according to the FB codebook for a specified $\Lcsi$ beams per rank up to the rank indicator, which is the maximum supported MIMO rank, $R$. A straightforward method for selecting the PMI would be to exhaustively try every possible combination of $\Lcsi$ beams from the codebook and select the best one (according to some metric). Assuming the feedback codebook is an oversampled DFT codebook, we can make use of efficient algorithms for quantizing the beamformed channel estimate into the PMI. Although not presented here, codebooks can also be used for SU-MIMO feedback; we use this form of feedback when comparing against the SU-MIMO results in Section \ref{sec: sim_results}.
        

        First, we define the oversampled DFT codebook $\Bdft$ beginning with the (non-oversampled) DFT codebook for a uniform linear array $\vect{U}_N$ defined in the same way as e.g. \cite{HeathOverviewSP4mmWave2016}. Then we can define a diagonal oversampling matrix $\vect{D}_{O, N}$ for an oversampling $O$ and $N$ elements as 
        \begin{align}
            \vect{D}_{O, N} = \text{diag}\left(1,\ \exp{\left(-\frac{\mathsf{j} 2 \pi}{O N}\right)}, ...,  \exp{\left(-\frac{\mathsf{j} 2\pi (N-1)}{O N}\right)}\right).
        \end{align}
        Then, the oversampled DFT codebook for a one-dimensional array can be concisely written as
        \begin{align}
            \vect{B}_{O, N} = [\vect{U}_N, \vect{D}_{O, N} \vect{U}_N, \vect{D}_{O, N}^2 \vect{U}_N..., \vect{D}_{O, N}^{O-1} \vect{U}_N] \in \complex^{N \times O N}.
        \end{align}
        Finally, the UPA oversampling codebook can be defined as the kronecker product of the $\Nx$ and $\Ny$ dimensions
        \begin{align}
            \Bdft = \vect{B}_{\Oh, \Nx} \otimes \vect{B}_{\Ov, \Ny} \in \complex^{\Nx\Ny \times \Oh\Nx\Ov\Ny}. \label{eqn: Bdft}
        \end{align}
        With the FB codebook defined \eqref{eqn: Bdft}, the PMI selection procedure can begin. The UE first calculates the inner product for all beams from the oversampled codebook $\Bdft$ and selects the first beam (and by association the oversampling basis) as
    \begin{align}
        \vect{C} &= \HFest \Bdft \ \in \complex^{\Nr \times \Nx \Oh \Ny \Ov} \\
       Q_0 &= \argmax |{\vect{C}}|
    \end{align}
    where the argmax is determined from the absolute value of all elements of $\vect{C}$ and the resulting $Q_0$ is an index corresponding to oversampling powers $Q_{0,X}$ and $Q_{0,Y}$. Because (non-oversampled) DFT beams are orthogonal, the PMI can be calculated iteratively by selecting the first beam from the entire oversampled codebook $\Bdft$, reducing the codebook to just the orthogonal subset, and selecting the beams from the subset. The orthogonal subset can be understood as the kronecker product of the oversampling-shifted DFT codebooks 
    \begin{align}
    \vect{B}_{\text{ortho}, Q_0} = \vect{D}_{\Oh, \Nx}^{Q_{0,X}} \vect{U}_{\Nx} \otimes \vect{D}_{\Oh, \Nx}^{Q_{0,Y}} \vect{U}_{\Ny}
    \end{align}
    where $Q_{0,X}$ and $Q_{0,Y}$ are the oversampling powers in the X and Y directions of the codebook index $Q_0$.
    With the orthogonal subset, the next step is to select the $\Lcsi$ beam indices $\q{} \in \mathbb{Z}_+^{\Nr \times \Lcsi}$ and combining coefficients $\Cq \in \complex^{\Nr \times \Lcsi}$ 
    \begin{align}
        \vect{c}_{\text{ortho}} &= \HFest_{\nr} \vect{B}_{\text{ortho}, Q_0} \ \in \complex^{\Nx \Ny} \label{eqn: scores}\\ 
        \q{\nr} &= \kargmax{(\Lcsi)} |{\vect{\vect{c}_{\text{ortho}}}}| \quad \forall \nr \label{eqn: kargmax} \\ 
        \Cq_{\nr} &= \vect{c}_{\text{ortho}}\left[\q{{\nr}}\right]  \quad \forall \nr \label{eqn: Cq} \\
        \Cq_{\nr} &= \frac{\Cq_{\nr}}{\Cq_{\nr}[\argmax{|\Cq_{\nr}|}]} \quad \forall \nr.\label{eqn: normalize}
    \end{align}
    First, the orthogonal beams are ``evaluated'' in \eqref{eqn: scores}, then we make use of the $\kargmax{k}$ operation defined as the $k$ indices for the $k$ largest values in \eqref{eqn: kargmax} to select the $\Lcsi$ beam indices. Finally, the combining coefficients are determined as the complex values of the beam products \eqref{eqn: Cq} and normalized according to the largest coefficient \eqref{eqn: normalize}. The oversampling basis $Q_{0,X}, Q_{0,Y}$ is reported back in the PMI so that the overall beam selections can be reconstructed at the BS. 

        
        The CSI report contains the CSI-RS indicator (CRI) which is the strong CSI-RS beamforming codeword, the rank of the PMI, the RSRP or SNR, and the PMI feedback beam indices $\q{}$, as well as $4$ bit amplitude and $8$-PSK phase values. The amplitude and phase values come from $\Cq$ which are the corresponding complex values for each $[n_r, \ell]$ feedback beam. There is an additional overhead reduction step in the specification where instead of indexing each element of $\q{}$ from the entire codebook, the set of indices is translated to one of the ${\Nx\Ny \choose \Lcsi}$ combinations and one index corresponding to the oversampling factors of the strongest index $(Q_{0, X}, Q_{0, Y})$ is used. In this paper, we will ignore the amplitude and phase quantization (assuming perfect amplitude and phase knowledge) to focus on the effects of the codebooks, SNR, and feedback. The CSI report can also contain both wideband and narrowband components for a degree of frequency selective precoding. The number of narrowband components is small, at most $\bwp=8$ in FR1, to mitigate the growing feedback and complexity. All of the previous and subsequent steps are defined for $\bwp=1$, but extending to frequency selective feedback and precoding is straightforward. We investigate how the amount of feedback ($\Lcsi, \bwp$) impacts performance in Section \ref{sec: sim_results}.

    \subsection{MU-MIMO data transmission}
        After the CSI-RS period, the base station must determine the precoder that serves a group of users, often with the goal of maximizing the sum rate or proportional fair rate. Until this stage, all of the downlink transmissions have been multi-cast in the sense that the same pilots and data are intended for all users. Therefore, there is no interference assuming a well-designed OFDM cyclic prefix and subcarrier spacing is used with accurate synchronization. In contrast, the downlink data transmission carries different data for each user, resulting in interference between data streams. The remainder of this section outlines the channel reconstruction process, precoder selection, and achievable spectral efficiency in the downlink channel. 
        
        First, because the feedback is type-II, the BS can reconstruct the PMI (corresponding to the beamformed channel estimate here) with the corresponding complex scaling $\Cq$ as 
        \begin{equation}
            \widehat{\widehat{\vect{HF}}}[n_r] = \sum_{\ell=0}^{\Lcsi-1} \Cq[n_r, \ell] \Bdft[:, \q{n_r, \ell}] \quad \forall\ \nr.
        \end{equation}
        Then, assuming the beamformers were well-chosen such that $\Fcsi$ is right-invertible, channel reconstruction can be performed successfully. We reintroduce the UE $(\ue)$ notation here to allow for aggregating all user channels in subsequent steps by 
        \begin{align}
            \Hbsu &=  \widehat{\widehat{\vect{H}_{\ue}\vect{F}}} \F_{\text{CSI-RS}}^{\dagger} \label{eqn: HFinv}.
        \end{align}
        In the case of DFT beams, this process simply becomes an IDFT operation \cite{Morozov2016}, which is significantly more efficient than the matrix inverse of an $\Np \times \Np$ matrix. The efficiency at this step is critical to quickly using the feedback to serve users and is one reason we will tailor our algorithm to integrate with a DFT codebook for CSI-RS. Aggregating the results for each user, the base station obtains the downlink channel estimate $\Hbs \in \complex^{U \times \Nr \times \Np}$
        \begin{align}
            \Hbs &= \left[\Hbs_0, \Hbs_1, ..., \Hbs_{U-1}\right]. \label{eqn: chan_agg}
        \end{align}
        The base station can design the precoders for the set of estimated channels using any form of precoding. To reasonably evaluate the system we will assume a regularized zero forcing precoder is used with regularization determined to minimize the signal-to-leakage noise ratio \cite[section 9.9]{HeathLozano2018}. The RZF precoder is built up for each user as
          \begin{align}
            \vect{F}_{\ue} = \sqrt{\frac{\Nt}{U}} \frac{(\sum_{i=0}^{U-1} \Hbs_i^*\Hbs_i + U \Nt\mathbb{E}[\vect{N}_{t, k, u}^2])^{-1} \Hbsu^*}{\norm{  (\sum_{i=0}^{U-1} \Hbs_i^*\Hbs_i + U \Nt \mathbb{E}[\vect{N}_{t, k, u}^2])^{-1} \Hbsu^*}}.
        \end{align}
        The noise factor, $\mathbb{E}[\vect{N}_{t, k, u}^2]$ can be estimated based on the RSRP/SNR reported and the bandwidth of the downlink signal.
        The UE will also perform a combining strategy based on the previously determined rank. In this work, we will assume an LMMSE receiver is used to maximize the signal-to-interference noise ratio (SINR) for any precoder, using the embedded DMRS to determine the combined channel and precoder. The resulting SINR at resource element ($t, k$) for a given user $\ue$ and layer $r$ with equal power allocation per user is obtained as
        \begin{align} 
            \vect{I}_{t, k, u} &= \bigg(\sum_{i=0}^{U-1} \Htkstack{u} \vect{F}_i \vect{F}_i^*\Htkstack{u}^* + U \Nt \mathbb{E}[\vect{N}_{t, k, u}^2] \bigg)^{-1} \\
            \sinr_{t, k, \ue, r} &= \frac{1}{K}\frac{\vect{F}^*_{\ue}[:, r] \Htkstack{u}^* \vect{I}_{t, k, u} \Htkstack{u} \vect{F}_{\ue}[:, r]}  
            {1 -\vect{F}^*_{\ue}[:, r] \Htkstack{u}^* \vect{I}_{t, k, u} \Htkstack{u} \vect{F}_{\ue}[:, r]}. \label{eqn: sinr}
        \end{align}
        
        While the SINR expression \eqref{eqn: sinr} appears complicated, it is a simplified ratio between the signal power for the data stream $(\ue, r)$ versus the interference power of all other streams. The SINR at this step is not calculated in a physical system, but it is necessary for evaluating the achievable spectral efficiency.
        
        Finally, the most critical metric in the wireless network is the sum SE or sum rate when applied to a specific bandwidth. A fairness constraint can also be applied, however, we have not specified any form of scheduling so we are most interested in maximizing the sum spectral efficiency or rate. Assuming Gaussian signaling and treating interference as Gaussian noise, the achievable spectral efficiency, $\text{SE}_{\ue, t}$, is 
        \begin{align}
            \text{SE}_{\ue, t} = \sum_{k=0}^{K-1} \sum_{r=0}^{R-1}{\log_2(1 +\sinr_{\ue, t, k, r}}).
        \end{align}
        In the final post-processing, we consider the effective sum SE, which accounts for the overhead due to beam training by removing the corresponding time/frequency resources due to training and feedback of the beam management system $(T_{\text{BM}}, K_{\text{BM}})$ from the spectral efficiency calculation
        \begin{align}
            \text{Eff-SSE} = \sum^{U-1}_{\ue=0} \sum_{t \notin T_{\text{BM}}} \sum_{k\notin K_{\text{BM}}} \sum_{r=0}^{R-1} \log_2 (1 + \text{SINR}_{\ue, t, k, r}).
        \end{align}
        With this goal and the beam management framework in mind, we can now define the beamspace processing and machine learning codebook design algorithm.
    
	\begin{figure}[!t]
	    \centering
	    \includegraphics[width=3.25in]{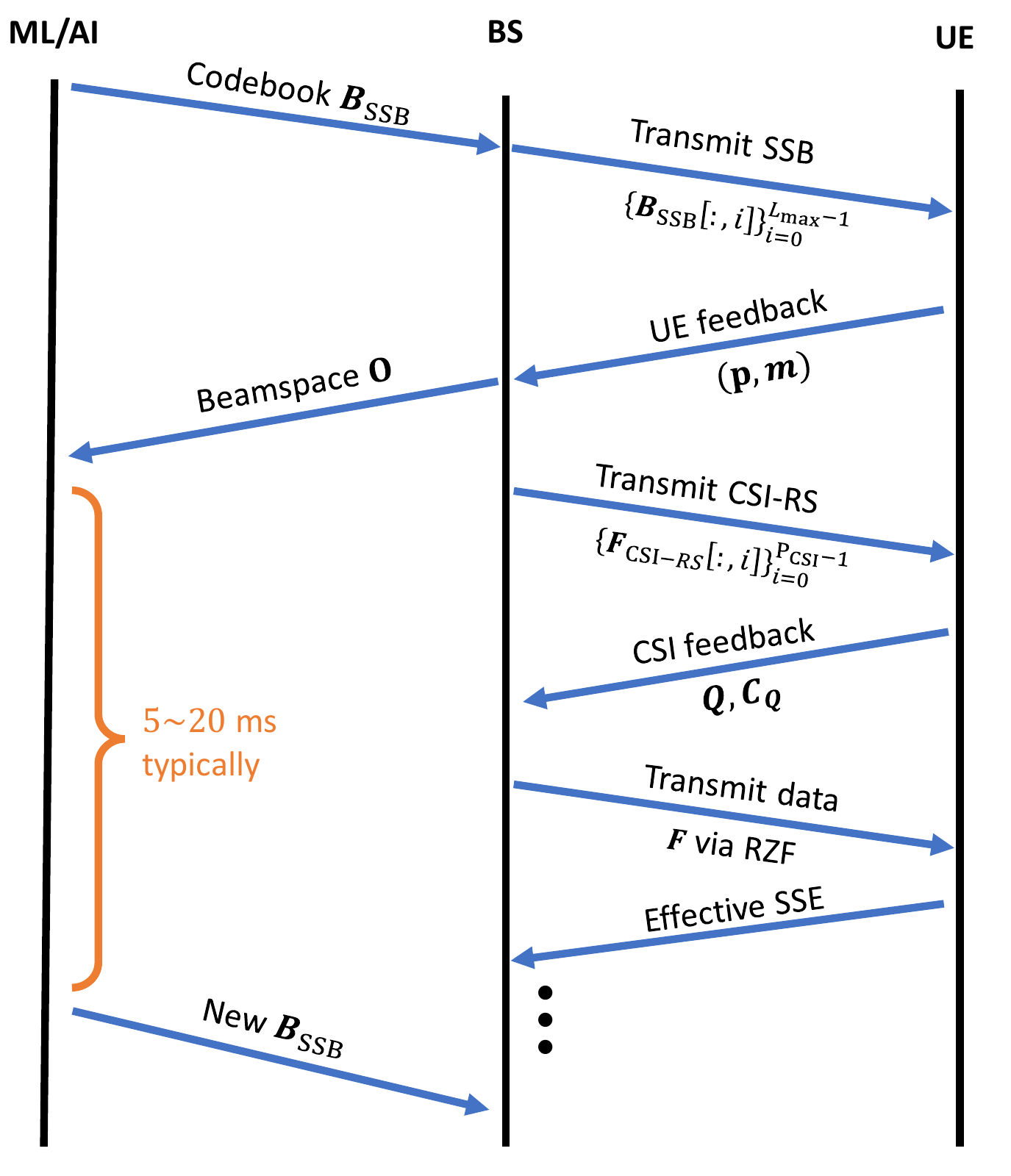}
	    \caption{A timing diagram of the process for codebook learning and evaluation. The AI/ML engine provides new codebooks at each SSB interval for the base station. At each CSI-RS period, the BS transmits pilots using the CSI-RS codebook, which is determined based on the SSB codebook and feedback. The CSI-RS feedback is used to determine the precoder used for data transmission.}
	    \label{fig: problem_statement}
	\end{figure}

    \begin{figure}[!t]
	    \centering
	    \includegraphics[width=6.25in]{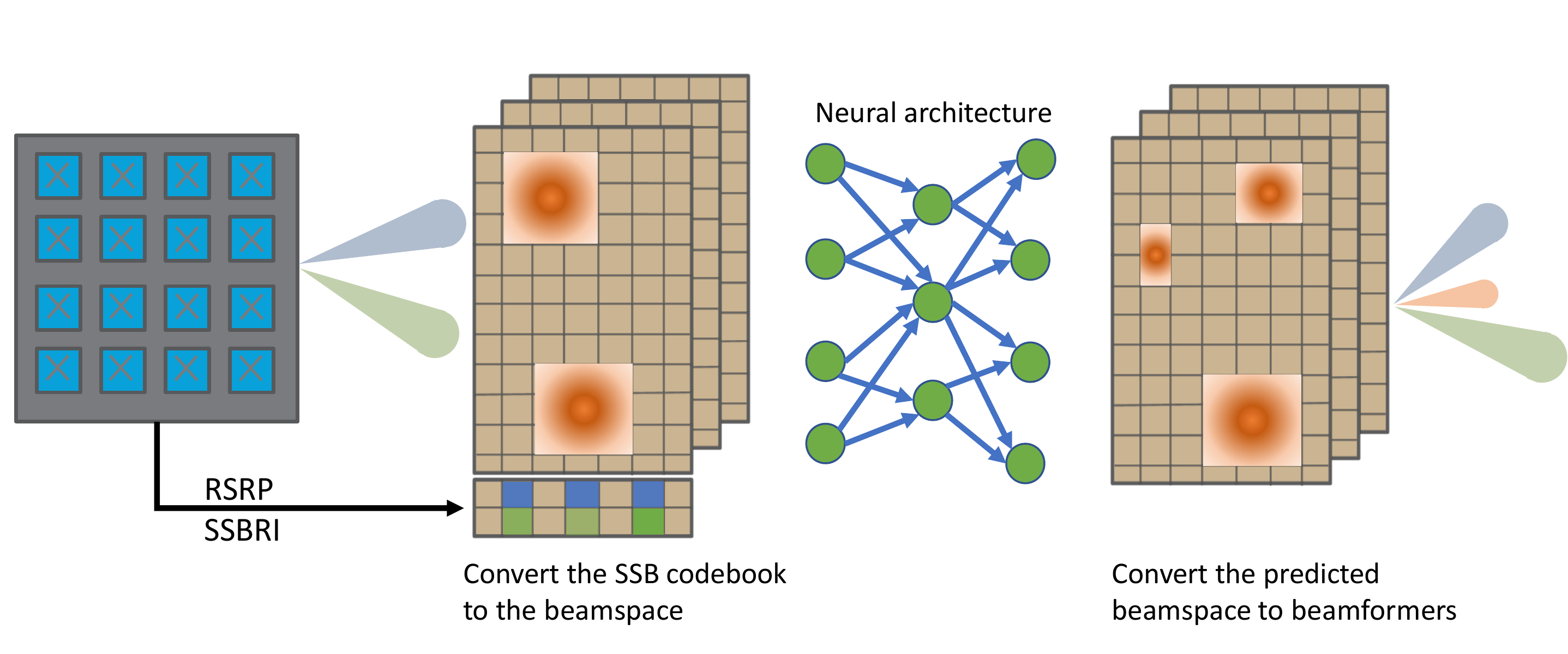}
	    \caption{A depiction of the BSC pre-processing, neural architecture, and post-processing. The feedback is converted to an angular representation (beamspace) and the strongest directions are identified based on the feedback reported. The neural network jointly processes the SSB feedback and beamspace representation to produce a new beamspace codebook that is then converted back to complex beamforming coefficients. This unifies the various codebooks with a consistent input dimensionality even with varying numbers of users or even antenna sizes.}
	    \label{fig: beamspace}
	\end{figure} 
	
\section{Beamspace-Codex} \label{sec: algorithm}
    In this section, we present the Beamspace-Codex which is a neural network architecture and processing setup that is used to generate the SSB codebooks, with considerations for joint SSB-CSI-RS codebook design left to future work. We start by introducing the beamspace observation that provides a consistent basis for a learning algorithm using dynamic codebooks. We then present the supervised learning formulation that enables convergent and dynamic codebooks. Finally, we define the neural network architecture that enables learning the underlying relationship between the beamspace observation and the SSB codebook. The full methodology--processing, architecture, training, and evaluation--is critical to integrating wireless domain knowledge and physical structure into the learning algorithm.
    
    \topic{One of the primary steps in designing an appropriate neural network is formulating the input-output relationship to be conducive and consistent for learning.} There are two problems with neural networks within the context of 5G beam management. First, the SSBRI feedback corresponds to a beamformer from a codebook that was used previously, which does not provide a consistent representation or meaning with dynamic codebooks. Second, the number of users is not constant within a cell, so the size of the feedback and data dimension changes over time. Although there are specific ways to overcome dynamic data sizes, it often requires sacrificing hardware optimizations and impedes inference and training times.
    
    We propose transforming the codebook into the beamspace domain. The beamspace is an angular representation of a beamformer corresponding to the array factor evaluated over a range of directions \cite{Sayeed2002Beamspace}. This ensures that regardless of the specific beamforming codeword, or even physical antenna dimension, the corresponding input from a set of previous beamformers still represents a two-dimensional grid of projections. The beamspace conversion for $N_{x0}$ azimuth directions and $N_{y0}$ elevation direction is calculated as
    \begin{align}
        \boldsymbol{\theta}_{N_m} &= \frac{1}{\pi}[0, 1, ..., N_m-1]^T\\
        \vect{U}_{N_s, N_m} &\triangleq [\vect{a}_{N_s}(\boldsymbol{\theta}_0), ... \vect{a}_{N_s}(\boldsymbol{\theta}_{N_m-1})] \in \complex^{N_s \times N_m} \label{eqn: ang_matrix}\\
        \Fssb^{(\Nx, \Ny)}[i] &= [\Fssb[i, 0:\Nx]^T, \Fssb[i, \Nx:2\Nx]^T, ..., \Fssb[i, (\Nx-1)\Ny:\Nx \Ny]^T] \label{eqn: reshape}\\
        \vect{O}_{\text{BSC}}[i] &= \vect{U}_{N_{x0}, \Nx} \Fssb^{(\Nx, \Ny)}[i] \vect{U}_{\Ny, N_{y0}} \quad \forall i \in \{1, 2, ... \Lmax \}.  \label{eqn: beamspace}
    \end{align}
    The conversion starts by generating an angular matrix from \eqref{eqn: ang_matrix} for the azimuth directions $\vect{U}_{N_{x0}, \Nx}$ and elevation directions $\vect{U}_{\Ny, N_{y0}}$ as a series of array responses. Then,
    the codebook must be reshaped from a vector of size $\Nt$ to the planar dimensions $N_X \times N_Y$ in \eqref{eqn: reshape} before the beamspace conversion in \eqref{eqn: beamspace} that produces the beamspace observation, $\vect{O}_{\text{BSC}}$. In addition to the angular representation, the input is also concatenated with the feedback corresponding to the number of users reporting each beam and the sum RSRP after min-max normalization. Note that the beamspace conversion is a reversible operation, so we also train the network to predict the beamspace of the desired output, rather than direct beamforming coefficients. The predicted beamformers are obtained in post-processing, and the computational complexity can be controlled by changing the observation sizes $N_{x0}$ and $N_{y0}$. With the pre/post-processing defined, we can now define the supervised learning environment and desired outputs used for training.
    
    \topic{Supervised learning provides many advantages over unsupervised or reinforcement learning when there is a desired output compared to just a black-box metric.} 
    The task of determining beamforming vectors that maximize the power to a set of users has an optimal solution if CSI is available via singular value decomposition (SVD) beamforming. To that end, we assume a dataset of channels has been built up or simulated to allow for offline training, although our site-specific results in Section \ref{sec: sim_results} suggest that transferring a learned model between sites is still more effective than DFT codebooks. Then, the solution to maximize the power received for a  user is the right singular vectors (RSV) of the SVD of the channel \cite{Raghavan2016BFIA}. In the MU-MIMO OFDM case, this means taking the RSVs corresponding to the $\Lmax$ strongest singular values of all users. 
    Additionally, the timing of the beams is important and each beam has a specified set of resources on which to be transmitted. We must sacrifice the time relationship of the data, however, to build a true supervised setting. In particular, we do not use information from the past SSB feedback beyond the previous step. This limitation removes the capability of using historical data as inputs but significantly improves convergence compared to a reinforcement learning setting. The process is restricted to a two-step relationship where we include the previous codebook and feedback to help design the next codebook. We now have a supervised framework and pre/post-processing step that converts the previous codebook and RSV codebook to the beamspace domain. The final component of the Beamspace-Codex is the neural architecture that facilitates efficient learning.

    Determining the best neural network architecture is often challenging even with hyperparameter tuning and an intuitive understanding of the problem. Using the beamspace conversion, the problem can roughly be seen as a translation task from the initial beamspace to the RSV beamspace. It is not necessarily obvious how the limited feedback information could be used to recover the RSV codebook. There is some underlying information available, however, by the selection of one beam over $\Lmax-1$ other options. Additionally, the environment and user densities follow a pattern that is also possible to exploit. At the same time, the inputs and outputs are similar to an image of the angular directions of transmitted power. Therefore, we considered traditional architectures (fully-connected networks), image-based architectures (convolutional and diffusion \cite{Ho2020Diffusion} networks), and translation-based architectures (transformers \cite{Liu2021SWIN} and autoencoders). Although the problem has many similarities with both vision and translation problems, state-of-the-art architectures were not necessarily best under reasonable computational limits i.e. $24$GB of VRAM and $5$ms inference time. Throughout our testing, the fully connected architecture showed the strongest results with an average of $1$dB higher average RSRP than other architectures, while also being the most computationally efficient one. 
    
    We propose a multi-layer fully connected network with significant regularization through dropout comprised of $6$ densely connected layers with parameters and sizes determined through Bayesian hyperparameter tuning and defined in Table \ref{tb: model}.
    The model is trained using an Adam optimizer \cite{kingma2014adam} with early stopping and cosine learning rate decay. Training is performed with a $550,000$ sample training dataset following the setup in Section \ref{sec: simulation_results} with validation and test sets each from an unseen portion of the data corresponding to $160,000$ and $80,000$ samples respectively. Gradients are determined by the cosine distance ($d_{\text{cos}}(\vect{a}, \vect{b}) = 1 - \vect{a}^*\vect{b} / (\norm{\vect{a}}\norm{\vect{b}})$) between the RSV beamspace and the neural network-predicted beamspace. The choice of the loss function is critical for performance as the same network architectures trained on mean squared error results in significantly worse ($>5$dB) performance due to focusing on the pixel-wise magnitude, rather than the relative magnitude or direction of the RSV beamspace. 
    
    {\rowcolors{2}{blue!10}{}
    \begin{table*}[!t]
        \caption{Beamspace-Codex network parameters}
        \label{tb: model}
        \centering
        \renewcommand*\arraystretch{1.15}
        \begin{tabular}{|c|c|c|c|}
        \hline
        \textbf{Layer}       & \textbf{Primary Parameter}   & \textbf{Activation Func.} & \textbf{Output Dimension}                \\ \hline
         Flatten             &                              &                              & ($2\Lmax (N_{x0}+2)(N_{y0}+2)$)           \\ \hline
         Fully Conn.         & 144 Neurons                  & ReLU                         & (144)                                    \\ \hline
         Dropout             & 0.2 Rate                     &                              & (144)                             \\ \hline
         Fully Conn.         & 1808 Neurons                  & ReLU                        & (1808)                             \\ \hline
         Dropout             & 0.4 Rate                     &                              & (1808)                             \\ \hline
         Fully Conn.         & 272 Neurons                  & ReLU                         & (272)                             \\ \hline
         Dropout             & 0.4 Rate                     &                              & (272)                             \\ \hline
         Fully Conn.         & 240 Neurons                 & ReLU                         & (240)                             \\ \hline
         Fully Conn.         & 80 Neurons                  & ReLU                         & (80)                             \\ \hline
         Fully Conn.         & ($2N_{x0} N_{y0}\Lmax $) Neurons                 &                          & ($N_{x0}, N_{y0}, 2\Lmax $)                             \\ \hline
         Reshape             &                              &                              & ($N_{x0}$, $N_{y0}$, $2\Lmax$)           \\ \hline
        \end{tabular} 
        \normalsize
    \end{table*}
    }

\section{Data generation setup} \label{sec: sim_results}
     We require an accurate channel simulator coupled with a flexible framework that retains the timing of the 5G beam training process to fairly evaluate the performance of ML-designed codebooks. 
     We integrate the channel generation from QuaDRiGa \cite{JaeckelQuadriga} with a post-processing suite that enables an initial access scenario, beamspace generation, codebook determination, and evaluation of the network performance after accounting for beam management overhead. \topic{QuaDRiGa generates channels through a stochastic process with additional features for spatial consistency, correlations, and spherical wave modeling.} In our simulations, we simulate $200$ users scattered over a base station sector site. The BS is equipped with an $\Nx=4$, $\Ny=8$ planar array using a 3GPP 3D antenna model with half-wavelength spacing. Each user has $\Nr=4$ antennas all tuned for a 3.5GHz carrier frequency. $10\%$ of users are given a vehicular mobility pattern such that they travel along a roadway through the center of the cell with speeds normally distributed with mean $25$m/s and variance $5$m/s. The remaining users are uniformly scattered within $450$m and travel in any direction with speeds uniformly drawn from an interval of $[0, 3]$m/s. Channel distributions follow a 3GPP urban macrocell environment 
     and channels are sampled over a bandwidth of $100$MHz every $1$ms for $2$s. A large number of users ($\approx200$) are simulated to build up a database that is spatially and environmentally consistent. In the final section of the results, we address how the model generalizes to a new setting with different environmental parameters and different roadway locations. 
     
     Once the channels are generated, post-processing is necessary to produce a realistic setting and build up the algorithm's datasets. The process starts by randomly selecting a number of active users (uniformly between $[4, 12]$) and gathering the channels and time slots from the channel set. Then randomly selecting $\Lmax$ DFT beams for the `prior' codebook and calculating the RSV codebook. The codebooks are converted to beamspace representations and stored in a dataset along with the SSB feedback from a subset of the active users where each active user is assigned an $80\%$ probability of its feedback being included. This represents an initial access channel where not all users are previously known and introduces an additional regularization term into our algorithm to prevent over-focusing on known users. It is assumed that UEs use antenna selection for the RSRP reception during the SSB process, digital combining for CSI-RS, and LMMSE combining during data transmission as defined in the system model.

	
    
\section{Simulation results} \label{sec: simulation_results}
    \topic{It is critical to extensively evaluate a neural network model to fairly represent and compare it against traditional methods.} In our previous work \cite{Dreifuerst2022SignalNet}, we have considered a learning threshold as a minimal condition for determining if meaningful relationships have been learned by the network. In realistic wireless channels, such settings are much harder to characterize, so we instead benchmark the performance against industry-standard DFT techniques and RSV codebooks assuming CSI were known perfectly. This helps to identify the upper-performance limits (with respect to maximum RSRP) as well as the minimal performance needed to justify moving beyond traditional codebooks. We then use our robust simulation framework to better understand how the various codebooks and type-II CSI feedback parameters affect network performance in realistic FR1 channels. 
    In the first result, the RSRP performance of the SSB process with various codebooks is plotted. Then the subsequent SNR achieved during CSI-RS is evaluated, where the CSI-RS beams are determined by decomposing the SSB beam into a subset of DFT beams with an equal proportion to the number of users selecting the beam during SSB. 

	\subsection{RSRP and SNR}
\begin{figure*}[!t]
    	    \centering
    	    \subfloat[SSB RSRP \label{fig: RSRP_beams}]{\includegraphics[width=3.25in]{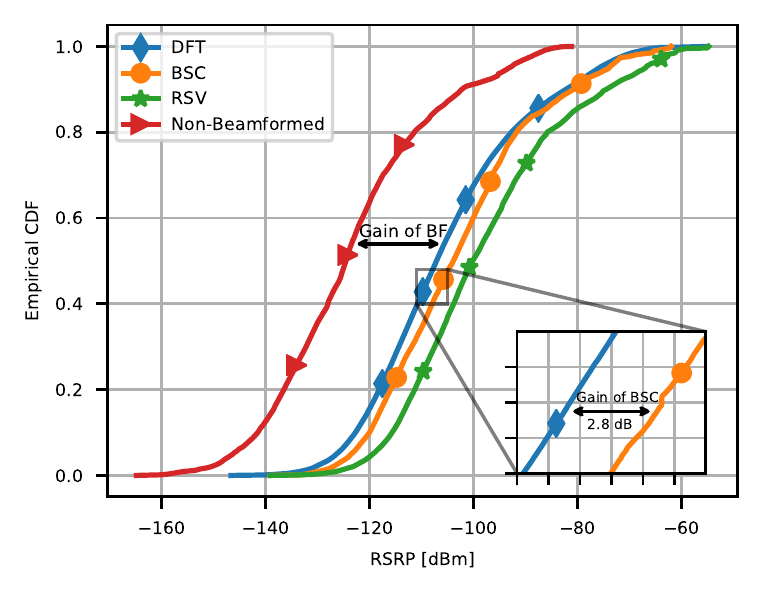}}
            \subfloat[CSI-RS SNR \label{fig: SNR_beams}]{\includegraphics[width=3.25in]{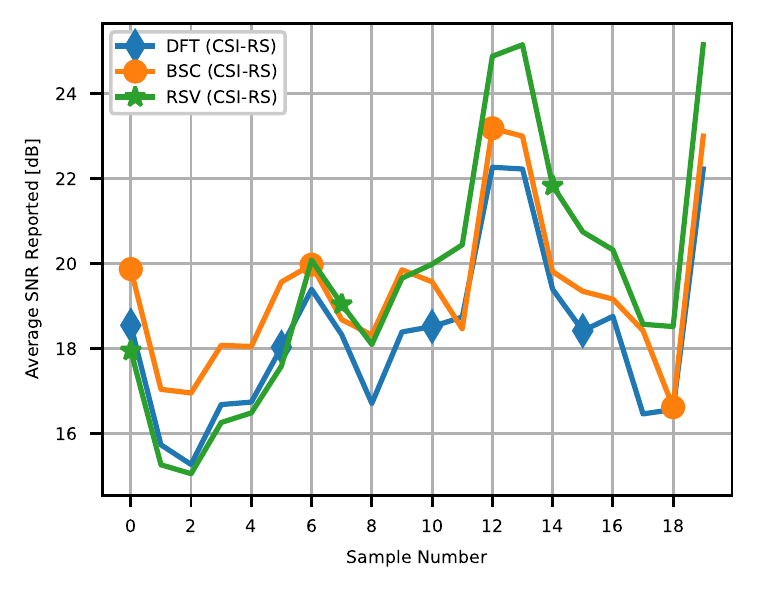}}
    	    \caption{Two plots of the codebook performance measuring (a) RSRP received during SSB and (b) SNR during CSI-RS reception. During the SSB stage, each of the codebooks is used directly and then in (b) the SSB codebooks are decomposed to form the active CSI-RS codebook subset. Non-beamformed transmission presents a significant loss in performance shown by the ``gain of BF'' arrow, while the learned BSC codebook shows gains of almost $3$dB over traditional DFT beams. While RSV beamforming is RSRP-maximizing during the SSB, the DFT decomposition of an RSV codebook is not necessarily SNR-maximizing due to multipath propagation.}
    	    
    	\end{figure*}

        The RSRP is a basic metric that can be used to directly demonstrate the ability of the neural network to output a valuable SSB codebook. Following the SSB process, we also highlight the performance that the dynamic codebook provides when selecting CSI-RS beams as a decomposition of the SSB codebook. While these two metrics--RSRP and SNR--do not ultimately characterize the network performance as a result of the codebook algorithm, they do provide meaningful insight into the performance of various codebook methods. Figure \ref{fig: RSRP_beams} shows the empirical cumulative distribution function (CDF) of the RSRP reported using various codebook/beamforming methods compared to our proposed BSC algorithm. Without any beamforming, more than $50\%$ of users would be below the minimum RSRP for signal detection, which is often around $-120$dBm. The SSB codebooks are decomposed into CSI-RS codebooks in Figure \ref{fig: SNR_beams}, where we see the RSV codebook is \textit{not} consistently the best SSB codebook when directly decomposed into CSI-RS codebooks, although the decomposed BSC is not universally better than decomposed DFT codebooks either. We can understand this as a result of the rich scattering environment that causes the RSV to not effectively decompose into a small number of incoherent DFT components. While RSV beamformers maximize the received power for a user, this maximization is based on the signal coherently combining and is not well-represented by a narrow DFT beam. From these results, we can see that the BSC algorithm is advantageous compared to traditional methods, but we also see that restricting the CSI-RS codebook to oversampled DFT beams is a significant limitation.
	
    \subsection{CSI-RS reporting}
        In this subsection, we provide an in-depth evaluation of the CSI-RS process without BSC codebooks. We evaluate performance based on the effective downlink SE after overhead due to beam training, uplink feedback, and an additional signaling overhead factor of $10\%$. Uplink feedback is assumed to be the only resource allocated for uplink transmission, and we assume a worst-case scenario where users transmit with MCS0 from Table 5.1.3.1-1 \cite{3gppTS38.214} with users allocated at the resource block level to account for additional overhead. User selection during the data phase is performed based only on the knowledge at the BS by first estimating the sum spectral efficiency achieved with the imperfect CSI for all users with $\{1, 2\}$ data layers. The best estimated SE combination is selected and used. Such a format is naturally suboptimal, even with perfect CSI, due to a lack of rank and power adaption. User selection is not the focus of this investigation, so we only use an exhaustive search over the reconstructed channel knowledge available at the BS to provide a basic but reasonable scheduling algorithm. The SE is also compared with rank-1 non-PMI beamforming where RZF is not employed, the channel is not estimated, and the only feedback is obtained from the CRI, with user selection and precoder determination simply being the CSI-RS beamforming codewords with the strongest reported user RSRP. This is often an envisioned beam management strategy for sparse mmWave channels with short, low-latency feedback, but is not assumed to be effective in sub-6GHz channels due to the rich scattering environment.
        
\begin{figure*}[!t]
    	    \centering
    	    \subfloat[Codebook size $\Pcsi$ \label{fig: SE-vs-P}]{\includegraphics[width=3.25in]{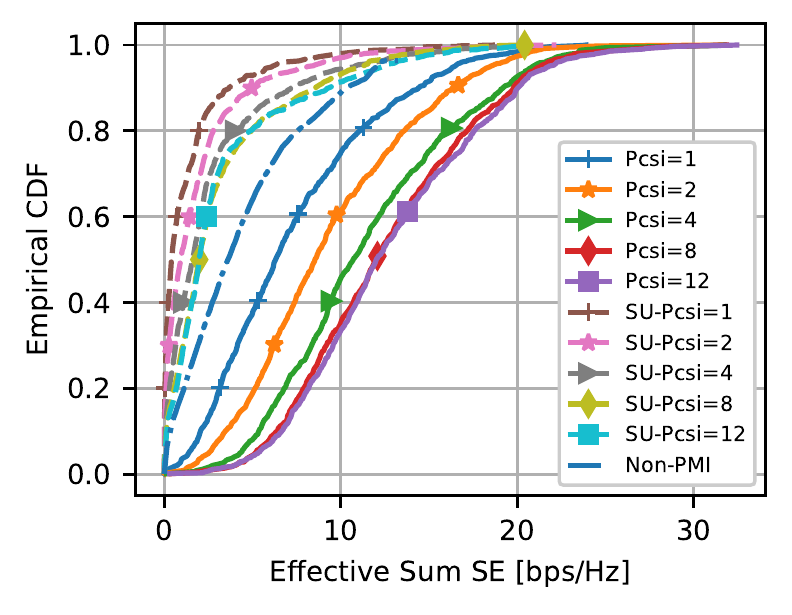}}
            \subfloat[Quantization paths $\Lcsi$ \label{fig: SE-vs-L}]{\includegraphics[width=3.25in]{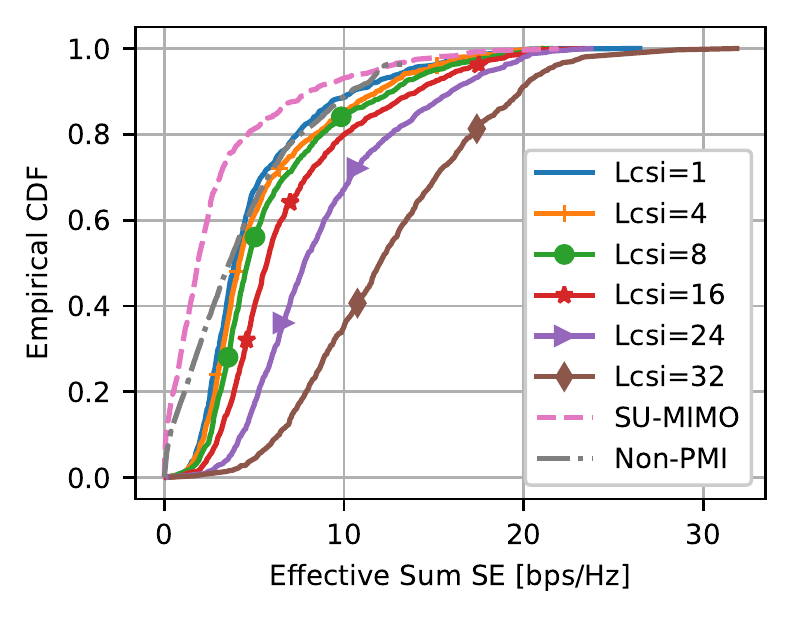}}
    	    \caption{Evaluation of the effects of active CSI-RS codebook size (a) and quantization using $\Lcsi$ feedback components (b). We can use the size of the codebook as a means of looking at how codebook optimization improves performance because the overhead with larger $\Pcsi$ is very small compared to the overhead of type-II feedback. SU-MIMO results using type-I PMI are shown by dashed lines while MU-MIMO results are shown by solid lines and Non-PMI refers to rank $1$ beamforming using the CRI. SU-MIMO does not include multipath feedback so only a single SU-MIMO result is shown in (b). Other parameters are set to $\Lcsi=32$ for (a), $\Pcsi=16$ for (b), $\bwp=1$, $\text{NRB}=24$. We can see that the loss from using $\Lcsi=4$ (3GPP Release 16 maximum support) compared to $\Lcsi=32$ is more detrimental than restricting the system to a single CSI-RS beamformer.}
    	\end{figure*}

        The size of the CSI-RS codebook is related to the channel estimation SNR because larger CSI-RS codebooks, made of orthogonal DFT beamformers, will result in equivalent or higher SNR compared to smaller codebooks with the same channels and feedback. We characterize the performance of larger or smaller CSI-RS codebooks in Figure \ref{fig: SE-vs-P}. Although the estimation SNR, and by extension performance before quantization, will always improve with a larger codebook, there is also a modest amount of overhead with larger $\Pcsi$ so that there is little spectral efficiency gain beyond $\Pcsi\ge 8$.
        
\begin{figure*}[!t]
    	    \centering
    	    \subfloat[BWP \label{fig: SE-vs-B}]{\includegraphics[width=3.25in]{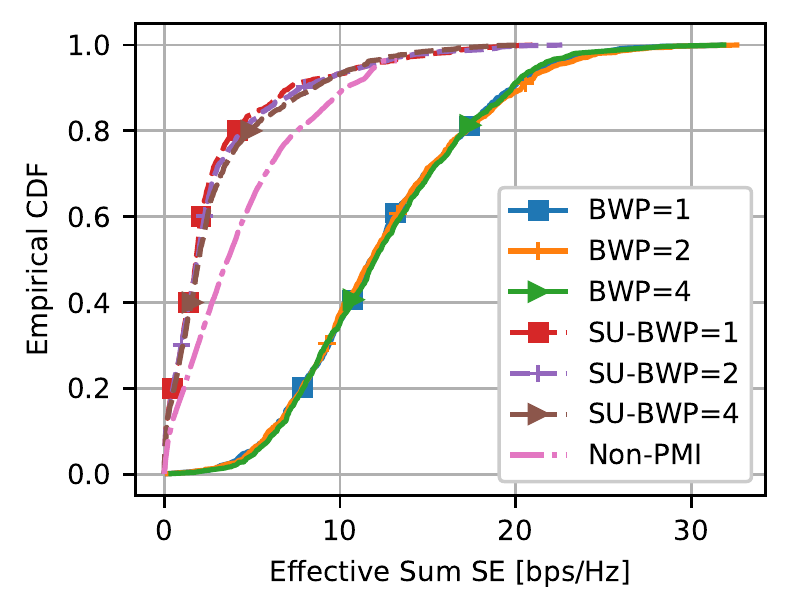}}
            \subfloat[NRB \label{fig: SE-vs-BWP}]{\includegraphics[width=3.25in]{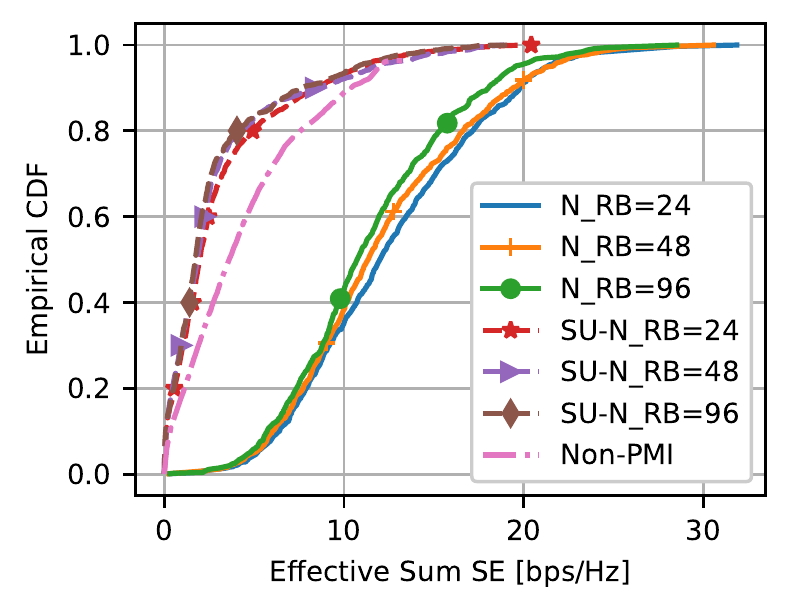}}
    	    \caption{A comparison of the effective spectral efficiency with different numbers of frequency selective precoders $\bwp$ and varying the number of resource blocks ($\text{NRB}$) per $\bwp$. It is assumed that $\Lcsi=32$, $\Pcsi=16$, $\text{NRB}=24 \times \bwp$ in (a) and $\bwp=1$ in (b). While increasing the number of resource blocks, therefore increasing the number of pilots, also increases the channel estimation performance, the rising feedback overhead causes the performance to decrease in (b). In comparison, the number of $\bwp$ increases the overhead linearly, yet the performance is essentially equivalent in (a). Therefore we see that frequency selective feedback and beamforming is not necessarily advantageous when overhead is accounted for.}
    	\end{figure*}

        In the next comparison, the feedback quantization is evaluated by modifying the parameter $\Lcsi$ in type-II formats. We can see from Fig. \ref{fig: SE-vs-L} that higher resolution feedback is beneficial, with the effective data rate exceeding $2$ times higher when the resolution is improved from $\Lcsi=1$ to $\Lcsi=32$. From there on, the other parameters CSI-RS resource block allocation ($\text{NRB}$) and the number of frequency selective resources ($\bwp$), are less influential compared to the quantization resolution. Most of the modest performance gains achieved with multiple frequency selective precoders and larger downlink CSI-RS resource usage are offset by the increasing overhead resulting in negligible or even reductive performance impacts. Joining the information from Figures \ref{fig: SE-vs-P}-\ref{fig: SE-vs-B}, a good balance of feedback and performance can be achieved with a setting of $\Lcsi=32$, $\bwp=1$, $\text{NRB}=24$, $\Pcsi=8$. It is important to note that the current 5G Release 16, even with enhanced type-II feedback, only supports $\Lcsi<=6$, which means most situations will only yield about $1.5$ times higher effective spectral efficiency relative to SU-MIMO, and essentially equivalent performance to the non-PMI beamforming. Perhaps even more surprising is that using a single CSI-RS beam ($\Pcsi=1$) has less degradation than using $\Lcsi=8$ feedback beams per rank, relative to the best performance. This suggests that feedback limitations are more likely to restrict MU-MIMO performance in sub-6GHz systems compared to codebook optimization or accurate channel estimation.

    \subsection{Site generalization}
\begin{figure}[!t]
    \centering
        \includegraphics[width=3.15in]{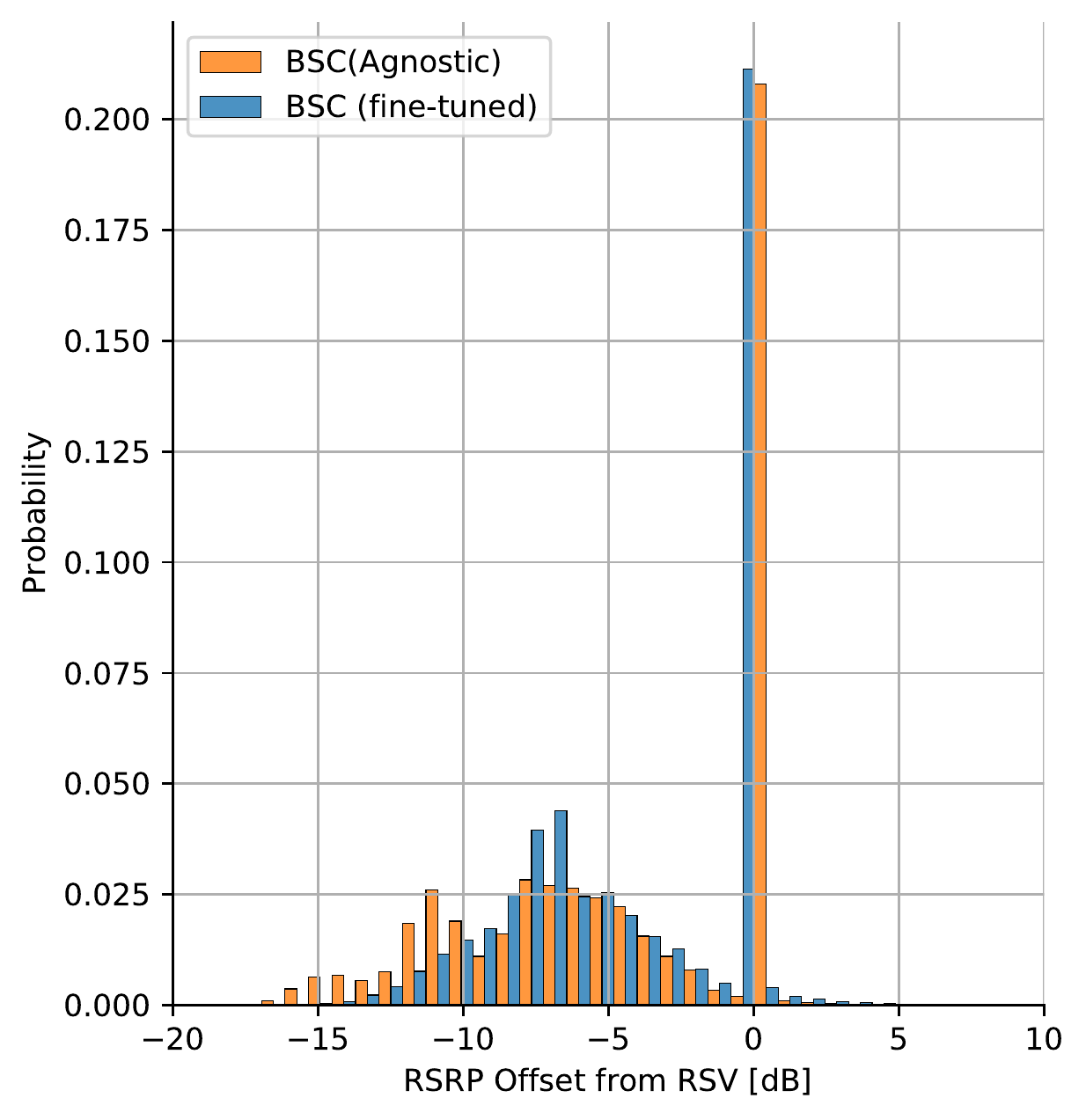}
        \caption{Histogram representing the performance delta between BSC and RSV codebooks in the agnostic setting and after fine-tuning. It can be seen that the worst-case differences are resolved and in some cases (when more users are present than SSB beams) the RSV codebook can be outperformed based on the higher proportion of the blue columns to the right of the orange columns.}
        \label{fig: site_specific}
    \end{figure}

	   Up to now, we have used a test set of data drawn from the same statistical environment as the training set. While a network operator will likely gather data for a specific site, here we examine the impact of testing on a new distribution of data. We simulate a new environment with completely different user distributions and mobility patterns. Now $50\%$ of users are vehicular with roadways defined in new locations and a new stochastic realization of the channel environment. In Figure \ref{fig: site_specific} the agnostic (orange) data is obtained without updating the model, whereas the fine-tuned data (blue) shows the results after we allow for $60$s of retraining on data from the new environment. Our time limit is arbitrary and could ultimately be orders of magnitude shorter than the timescale a network operator might use to update the model, but corresponds to approximately $1\%$ of the training time used in our setup for a balance of fine-tuning and generalization. It can be seen in Figure \ref{fig: site_specific} the performance in the new environment (agnostic) is significantly worse relative to the RSV. The performance loss is noticeably improved during fine-tuning, especially for the lowest $10\%$ of users.
 
\section{Conclusion}
    In this paper, we presented a novel framework for learning dynamic codebooks for sub-6GHz 5G NR. We set up a system with multiple UEs each equipped with a uniform linear array to receive from a base station with a fully digital planar array. We simulate realistic sub-6GHz channels using QuaDRiGa and build a post-processing framework for evaluating the system performance of SSB and CSI-RS codebooks and feedback. Using this framework we address two questions related to codebook performance in 5G: 1) Can machine learning design more effective SSB codebooks? 2) How should feedback be configured for MU-MIMO in sub-6GHz environments?
    
    We first proposed a codebook transformation based on beamspace projections that allowed for efficiently and consistently representing SSB codebooks in a format conducive to learning. This beamspace representation can circumvent challenges with dynamic user numbers, changing codebooks, and even deploying across different array sizes. Using the beamspace as a translation tool, we designed a neural architecture, Beamspace-Codex, to output new codebooks from SSB feedback. The proposed method directly integrates with current 5G standards and improves SSB RSRP by $2$-$3$dB on average. We also show that only $1\%$ of the training compute budget is needed for retraining in new environments to significantly improve the worst-case performance.
    Secondly, 
    we found that the PMI quantization resolution, determined by the number of feedback DFT beams, is a limiting factor in the performance of MU-MIMO in current standards. 

\bibliographystyle{IEEEtran}
\bibliography{IEEEabrv, RMD_all_refs, heath_refs_all} %
\end{document}